\documentstyle[floats,aps]{revtex}
\advance\topmargin 30pt
\def\subtilde#1{#1\llap{\lower2ex\hbox{$\widetilde{\hphantom{#1}}$}}}
\begin{document}
\input psfig
\pssilent
\title{Vassiliev invariants: a new framework for quantum gravity}

\author{Rodolfo Gambini$^{1}$\footnote{Associate member of ICTP.}, 
Jorge Griego$^1$, Jorge Pullin$^2$}
\address{1. Instituto de F\'{\i}sica, Facultad de Ciencias, 
Tristan Narvaja 1674, Montevideo, Uruguay}
\address{
2. Center for Gravitational Physics and Geometry, Department of
Physics,\\
The Pennsylvania State University, 
104 Davey Lab, University Park, PA 16802}
\date{Mar 5th, 1998}
\maketitle
\begin{abstract}
We show that Vassiliev invariants of knots, appropriately generalized
to the spin network context, are loop differentiable in spite of being
diffeomorphism invariant. This opens the possibility of defining
rigorously the constraints of quantum gravity as geometrical operators
acting on the space of Vassiliev invariants of spin nets. We show how
to explicitly realize the diffeomorphism constraint on this space and
present proposals for the construction of Hamiltonian constraints.
\end{abstract}

\pacs{4.30+x}

\vspace{-8.5cm} 
\begin{flushright}
\baselineskip=15pt
CGPG-98/3-1  \\
gr-qc/9803018\\
\end{flushright}
\vspace{6cm}

\section{Introduction}\label{sec:intro}

Quantum gravity presents a fundamental problem: it is a theory with an
infinite number of degrees of freedom, and yet it is ``topological''
in the sense of being invariant under diffeomorphisms. This makes the
description of the theory particularly troublesome. In particular, if
one recourses to structures of intrinsic topological nature, they tend
to have a discrete character that makes them incompatible with the
demands of a field theory with an infinite number of degrees of
freedom. To put this discussion in a more concrete setting, let us
concentrate on the canonical formulation of quantum gravity in terms
of the new variables introduced by Ashtekar. In terms of these
variables one can construct a representation of the quantum theory
purely in terms of loops (the loop representation), and a suitable
basis for describing quantum states is given by spin networks. That
is, quantum states are labelled by spin networks, multivalent graphs
embedded in three dimensional space, with a system of weights
associated to each side of the graph. If one imposes the
diffeomorphism constraint, one considers functions of the
diffeomorphism class of a spin network. There exists a precise sense
in which one can endow such a space with an inner product \cite{AsLe},
in terms of which spin network states are orthonormal.

In this context, there exists a proposal for the action of the
Hamiltonian constraint of quantum gravity due to Thiemann \cite{Th}.
In such a proposal the Hamiltonian constraint weighed by a lapse is
just given by a topological operator acting at the vertices of the
spin networks. The Hamiltonian constraint commutes with itself, as it
is expected it should happen in a diffeomorphism invariant
context. The construction also allows for the inclusion of matter in
the theory in a well defined regularized framework. It has been
observed \cite{LeMa}, however, that if one considers Thiemann's
Hamiltonian as acting on a larger space of functions (that are not
diffeomorphism invariant), the Hamiltonian constraint continues to
commute with itself. Although one can arrange the right hand side of
the commutator to also vanish, it appears that the price to pay is
tantamount to having a degenerate metric \cite{GaLeMaPu}. It also
appears that this result is rather generic, holding for many possible
detailed forms of the action of the Hamiltonian at vertices. Other
non-local proposals also seem to suffer of difficulties with the
constraint algebra \cite{Sm}. 

On the other hand over the last few years, a variety of formal
results have been obtained on a different space of states, that in
which the loop derivative is well defined. In particular, a proposal for
the Hamiltonian constraint of quantum gravity in terms of loop
derivatives exists, and it has been shown that at a formal level the
classical Poisson brackets are reproduced by the quantum theory
\cite{GaGaPu}. The main drawback of this space of states is
twofold: on the one hand there is the fact that the loop derivative
does not appear to exist on generic diffeomorphism invariant
states. This is due to the fact that such states change
discontinuously when one changes the loops and therefore one cannot
introduce a differential operator in loop space. Moreover, definitions
of the Hamiltonian in terms of the loop derivative have only been
attempted in the context of multiloops (not spin networks). In this
context, each type of loop and intersection has to be treated
individually, and therefore most results (for instance proofs that
certain states were annihilated by the constraints) were only of a
partial nature.

The purpose of this paper is to notice that there exist a set of
especially important knot invariants that appear to have the property
of being loop differentiable.  As a consequence, one can operate on
them in terms of the Hamiltonian and diffeomorphism constraints of
quantum gravity written in terms of loop derivatives in a more
systematic way. The invariants in question are the Vassiliev
invariants, which in turn are conjectured \cite{BaNa} 
to be complete enough to be
able to separate knots. Moreover, they have recently been generalized
to the spin network context \cite{GaGrPu88}. We will show explicitly that these
invariants are annihilated by the diffeomorphism constraint of quantum
gravity written in terms of loop derivatives. We will also start the
analysis of the action of the regularized Hamiltonian constraint of
the theory. All of this will be done in terms of spin networks, which
allows us to discuss in a unified way all types of intersections and
loops. 

The organization of this paper is as follows. In section II we discuss
the Vassiliev invariants and their loop differentiability. In section
III we discuss the action of the diffeomorphism constraint. In section
IV we analyze proposals for the Hamiltonian constraint. 

\section{Preliminaries}

\subsection{Quantum states}

The kind of states we will consider as candidates of interest for
physical states in quantum gravity has a long motivation in attempts
to try to show that there is a loop space representation counterpart
of the Chern--Simons state in quantum gravity (see \cite{GaPubook} for
references). To briefly summarize, one considers general relativity
written in terms of Ashtekar's new variables, which consist of a
canonical pair given by a set of densitized triads $\tilde{E}^a_i$ and
a (complex) $SU(2)$ connection $A_a^i$. If one considers a quantum
representation in which wavefunctions are functions of the connection
$\Psi[A]$, it is known that if one considers a wavefunction given by, 
\begin{equation}
\Psi[A] = \exp\left( { ik\over 4 \pi} \int {\rm Tr}(A\wedge \partial A
+{2 \over 3} A \wedge A \wedge A\right) =\exp\left(i k S_{CS}\right)
\end{equation}
is annihilated \cite{Ko,BrGaPunpb} by all the constraints of
quantum gravity with a cosmological constant $\Lambda= 24 \pi /k$.

One can now consider a loop representation, obtained by expanding the
states of the connection representation in terms of a basis of states
given the traces of the holonomies built with the connection (Wilson
loops). In terms of such a basis, the wavefunction we considered above
would be given by,
\begin{equation}
\Psi[\gamma] = \int D A \exp(i k S_{CS}) W(\gamma,A) 
\end{equation}
where $\gamma$ is a loop. This integral can be thought of in a
different context, as the expectation value of the Wilson loop in a
Chern--Simons theory. Such integrals have been the subject of a lot of
studies. It is known that the integral is related with a knot
polynomial that is a regular isotopy invariant (a framing dependent
knot invariant) known as the Kauffman bracket. It is also known that
all the framing information is concentrated in an overall factor
proportional to the exponential of the self-linking number of the
knot. If one removes this overall factor one is left with a polynomial
that is closely related to the Jones polynomial, which is ambient
isotopic invariant (a genuine invariant under diffeomorphisms)
\cite{Wi88}.  These polynomials appear evaluated for a particular
value of their variable, usually denoted as $q$. The particular value
is $q=\exp(2\pi i/k)$, with $k$ being the coupling constant of the
Chern--Simons theory. The resulting expressions are presented not as
polynomials but as infinite series in $1/k$. The coefficients of these
series are known to be Vassiliev invariants \cite{BaNa,Ba,KaBa}. Although
these results have never been derived using rigorous measure theory,
they have been used extensively to construct solutions to the
constraint equations. It was noted that one could write the
constraints of quantum gravity in the loop representation in terms of
differential operators in loop space involving the loop derivative,
and that one can (formally) act with these operators on states as the
one listed above. The end result of these manipulations was that one
can show that the exponential of the self-linking number is
annihilated by the Hamiltonian constraint with a cosmological constant
and that the second coefficient of the expansion in $1/k$ of the Jones
polynomial is annihilated by the vacuum (zero cosmological constant)
Hamiltonian constraint. This coefficient is a Vassiliev invariant.

All these results had several shortcomings. When one works in the loop
representation one needs to ensure that wavefunctions satisfy a series
of identities stemming from the fact that the loop basis consists of
traces of $SU(2)$ matrices. These identities are called Mandelstam
identities and in general imply very complex relations between the
values of functions of loops. Therefore if one wants to consider a
knot invariant as a possible candidate for a state of quantum gravity
in the loop representation, one needs to ensure that such invariant
satisfy the Mandelstam identities. Although this was the case for the
second coefficient we discussed above, it appeared as unlikely that
higher coefficients would satisfy the identities. Moreover, the proofs
that a certain coefficient was annihilated by the constraint were
performed for very special types of intersections (the Hamiltonian
constraint only acts nontrivially at intersections). Part of the
intention of this paper is to show that similar results are present in
the spin network context, where all types of intersections can be
treated in a unified fashion. A key to generalizing these results has
been the recent observation that ambient isotopic invariants (genuine
invariants under diffeomorphisms) can also be defined in terms of spin
networks \cite{GaGrPu88}.

\subsection{Spin networks}

Spin networks are constructed considering graphs that are embedded in
three dimensions. The graphs are multi-connected with intersections
that can be tri-valent or of higher valence. Each connecting line is
associated with a holonomy of the connection $A_a^i$ in a given
representation of the group in question (in our case, $SU(2)$,
representations are labelled by a (half)integer). One can construct a
generalization of the trace of the holonomy (Wilson loop) which we
call Wilson net, by considering the traces of the holonomies along the
Wilson nets joined by appropriate ``intertwiners'' at the
intersections. The intertwiners consist of invariant tensors in the
group. There are several possibilities of how to connect the
invariant tensors and the holonomies and this has led to different
conventions in the definition of the spin networks. The convention we
will choose will follow closely that of Witten and Martin
\cite{Wi89,Ma}, and differs of those of other authors in the field
\cite{RoSm,Th,KaLi}. 

Consider a multi-connected graph embedded in three dimensions. In order
to associate a holonomy to each line in the graph, we need to
associate an orientation to each line, so from now on we will consider
oriented graphs, with each line associated with a number which labels
the representation in which we are considering the holonomy along such
a line. The next task is to assign the intertwiners at each
intersection. Let us consider trivalent graphs. For graphs of higher
valence, one can choose a decomposition of the higher order
intersections into superpositions of graphs with trivalent ones
\cite{RoSm}. At each intersection one can have several possibilities,
depending on if the lines are all incoming, all outgoing or some are
ingoing and some outgoing. Let us consider the case of either all
ingoing or all outgoing lines. In those cases, there exist two ways of
contracting the holonomies with the $3jm$ symbol which intertwines
them, given that the latter is cyclic in its three entries. We will
label one of the possibilities ``$+$'' and the other ``$-$'' and we
will refer to them as the ``orientations'' of the vertex. For a
positive orientation we therefore have that the intertwiner is,
\begin{equation}
\raisebox{-10mm}{\psfig{figure=vi+.eps,height=20mm}} = 
\raisebox{-10mm}{\psfig{figure=vo+.eps,height=20mm}} = 
\left(\begin{array}{ccc} j_1&j_2&j_3\\m_1&m_2&m_3\end{array}\right),
\end{equation}
whereas for a negative orientation we have,
\begin{equation}
\raisebox{-10mm}{\psfig{figure=vi-.eps,height=20mm}} = 
\raisebox{-10mm}{\psfig{figure=vo-.eps,height=20mm}} = 
\left(\begin{array}{ccc} j_1&j_3&j_2\\m_1&m_3&m_2\end{array}\right) =
(-1)^{j_1+j_2+j_3} 
\left(\begin{array}{ccc} j_1&j_2&j_3\\m_1&m_2&m_3\end{array}\right).
\end{equation}
It is useful to introduce a tensor notation. In this notation we
denote the intertwiners as $V$ and use lower indices to denote
outgoing lines whereas upper indices denote ingoing lines. Therefore,
\begin{equation}
V_{j_1 m_1\,j_2 m_2\, j_3 m_3} = V^{j_1 m_1\,j_2 m_2\, j_3 m_3} =
\left(\begin{array}{ccc} j_1&j_2&j_3\\m_1&m_2&m_3\end{array}\right).
\end{equation}
Holonomies are also associated with tensorial objects, which we denote
by $U$. These objects are labelled by the appropriate representation
of the group $j$ and have two indices, the upper index is associated to the
point of departure and the lower index to the point of arrival, i.e., 
$U^{(j)}{}^n{}_n$. There is a metric tensor that allows to raise and
lower index,
\begin{equation}
\epsilon^{(j)}{}^{n\,n'} = \epsilon^{(j)}{}_{n\,n'} =
\left(\begin{array}{ccc} j&j&0\\n&n'&0\end{array}\right) = (-1)^{j-n} 
\delta^n_{-n'}.
\end{equation}
With this metric we can now obtain the intertwiners for trivalent
vertices with mixed ingoing/outgoing lines,
\begin{eqnarray}
\raisebox{-10mm}{\psfig{figure=vio.eps,height=20mm}} &=& 
V_{j_1 m_1\,j_2 m_2}{}^{j_3 m_3} = \epsilon^{(j_3)}{}^{m_3 m'_3} 
V_{j_1 m_1\,j_2 m_2\, j_3 m'_3}\\
\raisebox{-10mm}{\psfig{figure=voi.eps,height=20mm}} &=& 
V^{j_1 m_1\,j_2 m_2}{}_{j_3 m_3} = \epsilon^{(j_3)}{}_{{m'}_3 m_3} 
V^{j_1 m_1\,j_2 m_2\, j_3 m'_3}.
\end{eqnarray}
As an example of our conventions, let us consider, the evaluation of
the Wilson net for the so called ``theta diagram'' spin network,
\begin{equation}
\raisebox{-10mm}{\psfig{figure=theta+-.eps,height=20mm}} =
V_{j_1 n_1\, j_2 n_2 \, j_3 n_3} U^{(j_1)}{}^{n_1}{}_{m_1} 
U^{(j_2)}{}^{n_2}{}_{m_2} 
U^{(j_3)}{}^{n_3}{}_{m_3}  
V^{j_1 m_1\, j_2 m_2 \, j_3 m_3}.
\end{equation}
We have therefore provided a definition of the Wilson net for any spin
network. One simply constructs an invariant with holonomies and
invariant tensors as the ones described above. 

With the above definition, the Wilson net has certain properties. In
particular there is a dependence on the Wilson net on the orientation
of the vertices. What we will now do is modify the definition in order
to have an invariant that does not depend on the orientation of the
vertices. This will correspond to the normalization of the Wilson net
given by Witten and Martin. What we do is multiply the definition
introduced up to now times a factor given by,
\begin{equation}
V_\pm = \exp\left(\pm {i \pi\over 2} [j_1+j_2+j_3]\right) \sqrt[4]{
(2 j_1+1)(2 j_2+1) (2 j_3+1)}
\end{equation}
for each vertex in the spin network. With this factor, one can show
that the Wilson net defined is invariant under changes from  $+$ to $-$
type vertices. It is worthwhile noticing that in the work of Kauffman
and Lins \cite{KaLi} on invariants of spin networks the normalization
is different.

\subsection{Knot invariants for spin networks}

We wish to study invariants that represent the transform of the
Chern--Simons state into the spin network basis. That is, we are
interested in expressions of the form,
\begin{equation}
E(\Gamma,k) = \int D A \exp(i k S_{CS}) W(\Gamma,A) 
\end{equation}
where $\Gamma$ is a spin network and $W(\Gamma,A)$ is the Wilson net
we introduced in the previous subsection. This kind of integral has
been analyzed using monodromy techniques \cite{Wi89,Ma} and 
variational techniques \cite{GaPu,GaGrPu}. The result is a regular
isotopic invariant of spin networks. The techniques do not give a
unique answer for the invariant, but there are several possibilities,
tantamount to having several definitions for the measure $DA$. Each
possibility is uniquely characterized by prescribing a value for the
invariant on the so called theta-net. The choice of Witten and Martin
\cite{Wi89,Ma} is,
\begin{equation}
\Theta(j_1,j_2,j_3) = E\left(\raisebox{-12.5mm}
{\psfig{figure=thetaj1j2j3.eps,height=25mm}}\,,\,k\right) = 
\sqrt{\Delta_1 \Delta_2 \Delta_3}\label{theta},
\end{equation}
with,
\begin{equation}
\Delta_j =E\left(\raisebox{-5mm}{\psfig{figure=unknotj.eps,height=10mm}}
\,,\,k\right) = {q^{j+{1\over 2}} - q^{-j-{1\over 2}}\over q^{1\over
2} - q^{-{1\over 2}}} \label{delta},
\end{equation}
and as we mentioned above, $q=\exp(2 \pi i /k)$.

The definition of the invariant $E(\Gamma,k)$ is completely given in 
\cite{GaGrPu}, one needs to supplement the above definitions with
relations for recoupling and for crossings of the invariant. The
resulting invariant is regular isotopic, meaning that it is not
invariant under the elimination of twists. Specifically,
\begin{eqnarray}
E\left(\raisebox{-10mm}{\psfig{figure=y.eps,height=20mm}}\,,\,k\right)
&=& (-1)^{j_1+j_2+j_3}\exp\left(i\pi (h_1+h_2-h_3)\right) 
E\left(\raisebox{-10mm}{\psfig{figure=ytwist.eps,height=20mm}}\,,\,k\right),
\label{ytwist}\\
E\left(\raisebox{-10mm}{\psfig{figure=j.eps,height=20mm}}\hspace{-1cm},k\right)
&=& \exp\left(- 2 \pi i h_j\right)
E\left(\raisebox{-10mm}{\psfig{figure=jrtwist.eps,height=20mm}}
\hspace{-0.4cm},k\right)
=
\exp\left( 2 \pi i h_j\right)
E\left(\raisebox{-10mm}{\psfig{figure=jltwist.eps,height=20mm}}
\hspace{-0.4cm},k\right)\label{twist},
\end{eqnarray}
where $h_i \equiv {j_i (j_i+1)\over 2 k}$.

These moves correspond to diffeomorphisms, therefore the invariant
constructed is not invariant under diffeomorphisms. In other words,
the invariant is diffeomorphism invariant of ribbons and not of
ordinary loops. One needs a framing (a prescription for assigning a
ribbon to each loop) to have a well defined invariant. 

However, it was noted early on in the context of loops, and later 
in the context of spin networks \cite{GaGrPu88}, that the dependence on
framing can be concentrated on an overall factor. In order to isolate
this factor, one simply considers a power series expansion in $\kappa=
(2\pi i)/k$
and extracts the linear coefficient in $\kappa$. That coefficient,
exponentiated, is the  overall factor. This construction is discussed
in detail in \cite{GaGrPu88}, where it is shown that it indeed leads to
an ambient isotopic invariant, in other words, a genuine
diffeomorphism invariant function of loops. The resulting invariant
\footnote{Notice that since the definition of the theta-net involves
square roots of the variable, the invariant is no longer a polynomial,
as it was in the case of loops. It becomes a polynomial for the case
of colored links, i.e. spin networks without intersections.}
is then given by,
\begin{equation}
E(\Gamma,\kappa) = E(\Gamma,0) \exp\left(v_1(\Gamma) \kappa\right)
P(\Gamma,\kappa)
\end{equation}
where $P(\Gamma,\kappa)$ is a spin network generalization of the Jones
polynomial (it reduces to it when $\Gamma$ is a simple loop in the
fundamental representation) and $v_1$ is the first coefficient in the
expansion in powers of $\kappa$ (for a single loop it reduces to the
self-linking number). The factor $E(\Gamma,0)$, which corresponds to
the evaluation of the invariant for $\kappa=0$, contains information
about the coloring of the graph, and no information about the
embedding. It can be thought of as the evaluation of the Wilson net
for a flat connection.

It was shown by Alvarez and Labastida \cite{AlLa} 
in the case of $\Gamma$ being
simple loops, and later generalized for links, that the invariant
$E(\Gamma,\kappa)$ can be written as the exponential of a linear
combination of primitive Vassiliev invariants,
\begin{equation}
E(\Gamma,\kappa) = E(\Gamma,0) \exp\left(\sum_{i=1}^\infty
\sum_{j=1}^{d_i} \alpha_{ij}(\Gamma) r_{ij}(G) \kappa^i\right)
\end{equation}
where $\alpha_{ij}(\Gamma)$ are the primitive Vassiliev invariants,
which are only dependent on the embedding of the loop (they are
independent of the gauge group of the Chern--Simons theory).  $d_i$ is
the number of primitive Vassiliev invariants of order $i$, $G$ is the
gauge group (for our case $SU(2)$), and $r_{ij}$ are group-dependent
coefficients. We will assume that a similar structure appears for the
case of generic $SU(2)$ spin networks, specifically
\begin{equation}
E(\Gamma,\kappa) =
E(\Gamma,0) \exp\left(\sum_{i=1}^\infty
v_i(\Gamma) \kappa^i\right). \label{descomp}
\end{equation}

That is, we assume that the invariant is still given by the
exponentiation of a set of ``Vassiliev'' invariants for spin nets, but
we are not decomposing the expression in terms of ``primitive''
invariants, which is a structure that at present is not known in the
spin network context. The $v_i(\Gamma)'s$ are ambient isotopic for $i>1$. 

By constructing these invariants of spin networks, we have bypassed
one of the main obstacles that working with loops and links had in
this context: how to comply with the Mandelstam identities. The
invariants of spin nets we have constructed automatically take care of
this. An important missing element in the spin network context is the
idea of how ``generic'' is the set of $v_i(\Gamma)'s$. In the context
of loops it is conjectured that the primitive Vassiliev invariants are
enough to distinguish all knots. In the case of spin networks we are
not working with primitive invariants and therefore it is questionable
how generic the basis of invariants one is considering is. This is
important if one is making the case that these invariants are the
``arena'' where one is going to discuss quantum gravity. If a
decomposition in terms of primitive invariants of the exponential were
known, the techniques we will develop will still be
applicable. However, since up to present it is not known how to do
this decomposition we will work with the $v_i$'s. It is worthwhile
pointing out that the techniques we will introduce later in this paper
to operate with loop derivatives and diffeomorphism constraints are
geometrical in nature and not group dependent. Since it is known that
all Vassiliev invariants can be constructed from the Chern--Simons
integral with arbitrary groups, and our technique is not group
dependent, we therefore have a de-facto method to operate on all
Vassiliev invariants. For concreteness we will in the current paper
concentrate on the case of $SU(2)$. 

\section{Loop differentiability of the Vassiliev invariants}

\begin{figure}
\centerline{\psfig{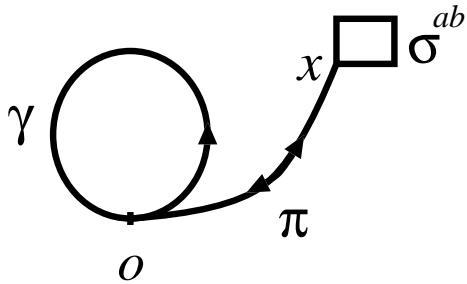}}
\caption{The loop that defines the loop derivative}
\label{loop}\end{figure}

The loop derivative \cite{GaPubook} is a differential operator in loop
space that arises by considering two loops that differ by an
infinitesimal element of area as ``close''. It acts on basepointed
objects (either loops or spin networks) by adding a path starting at
the basepoint up to a point $x$, where it introduces an infinitesimal
planar loop, and retraces back to the basepoint. The definition is,
\begin{equation}
(1+\sigma^{ab} \Delta_{ab}(\pi_o^x)) \Psi(\gamma) =
\Psi(\gamma\circ\delta\gamma) 
\end{equation}
where $\pi$ is a path from the origin to $x$, and the loop
$\gamma\circ\delta\gamma$ is shown in figure \ref{loop}
and $\sigma^{ab} = \delta u^{[a} \delta v^{b]}$ is the infinitesimal
element of area spanned by the two vectors $\vec{u}$ and $\vec{v}$
defining the parallelogram one adds at the end of the path $\pi$. For
a spin network, acting at a line on the network, the definition is
exactly the same as for loops, except that the path becomes a line of
the same valence as the line on which the basepoint lies, and the
infinitesimal loop is also of the same valence as the path. 

We now wish to apply the loop derivative to the invariants we
introduced in the previous section. A priori one expects that such a
quantity does not exist. In particular, for a generic knot invariant,
the loop derivative indeed does not exist. This is due to the fact
that knot invariants are discontinuous functions in the space of
loops. Loop derivatives change the topology of loops (for instance,
they can remove intersections \cite{BrGaPunpb}), and therefore the
limit defining the derivative is ill defined. What we will show here
is that due to the properties of the invariants of Chern--Simons under
deformations of the loops given by the skein relations, one can
introduce a reasonable definition of the loop derivative for these
kinds of invariants. It is similar to try to define the derivative of
a discontinuous functions by allowing the derivative to take value in
the distributions. We will analyze the consistency of this result with
the properties of the invariants. The strategy is as follows. The
invariants are defined as a functional integral of a Wilson net with a
weight function. The only dependence on the spin net is in the Wilson
net, so we will assume that the loop derivative of the invariant is
equal to the functional integral of the action of the loop derivative
on the Wilson net, with the appropriate weight factor,
\begin{equation}
\Delta_{ab}(\pi_o^x) E(\Gamma,\kappa) \equiv \int DA \exp(i k S_{CS}[A]) 
\Delta_{ab}(\pi_o^x) W(\Gamma,A).
\end{equation}
Here we have assumed that the limit defining the loop derivative and
the one defining the path integral are interchangeable.

\begin{figure}
\centerline{\psfig{figure=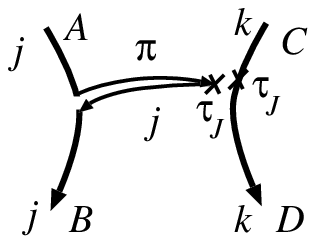,height=40mm}}
\caption{The loop derivative acting on a spin network}
\label{loopderspin}
\end{figure}

The starting point of the calculation is the action of the loop
derivative on a holonomy (in the $j$ representation of $SU(2)$) 
associated to an edge $e_A^B$ of the spin
network,  containing the basepoint $o$,
\begin{equation}
\Delta_{ab}(\pi_o^x) U^{(j)}(e_A^B)^m{}_n = 
[ U^{(j)}(e_A^o) U^{(j)}(\pi_o^x) F^{(j)}_{ab}(x)
U^{(j)}({\pi^{-1}}_o^x) U^{(j)}(e_o^B)]^m{}_n,
\end{equation}
and the fundamental relation satisfied by the Chern--Simons state,
\begin{equation}
F^{(j)}_{ab}(x) \exp({i k } S_{CS}[A]) = -{4 \pi i \over k}
\epsilon_{abc} {\delta \over \delta A_c^{(j)}(x)} \exp({i k } S_{CS}[A]).
\end{equation}
So the idea works exactly in the same way it did for ordinary loops
\cite{BrGaPunpb,GaPucmp}, the loop derivative acting on a holonomy
produces an $F_{ab}$, which can be re-expressed as a functional
derivative acting on the exponential of the Chern--Simons form. 
The last step is to perform a formal integral by parts of the
functional derivative and have it act on the holonomies of the spin
network that were left by the action of the loop derivative. The end
result is,
\begin{eqnarray}
\Delta_{ab}(\pi_o^x) E(\Gamma,\kappa) &=& -2 \kappa \sum_{e_C^D
\in \Gamma} \epsilon_{abc} \int_{e_C^D} dy^c \delta^3(x-y) \times\\
&&< \cdots U^{(j)}(e_A^o) U^{(j)}(\pi_o^x) \tau^J_{(j)}
U^{(j)}({\pi^{-1}}{}_o^x) U^{(j)}(e_o^B)\cdots U^{(k)}(e_C^y) 
\tau^J_{(k)} U^{(k)}(e_y^D) \cdots>_{CS}\nonumber
\end{eqnarray}
where $\tau^J_{(j)}$ are the $SU(2)$ generators \footnote{Our
convention for the $SU(2)$ generators is to take the Pauli matrices
divided by 2. This differs from other authors \cite{Th}.} in the $j$
representation, and the expectation value is assumed to be taken with
respect to the measure $DA \exp(i S_{CS})$ and the dots refer to the
fact that we just highlight the portion of the spin network where the
loop derivative acts. It is understood that the products of holonomies
continue until the net is closed and the appropriate traces are taken.
A pictorial representation of the quantity within the expectation
value is given in figure \ref{loopderspin}.

The above expression can be rearranged using the Fierz identity, 
\begin{equation}
\sum_{I=1}^3 [\tau_{(k)}^I]_s^t [\tau_{(j)}^I]_r^q = {1 \over 2} 
\sum_{l=|j-k|}^{j+k} (2 l +1) \rho_{l}(j,k)  
\raisebox{-10mm}{\psfig{figure=hvert.eps,height=20mm}},
\end{equation}
where $\rho_{l}= l(l+1)-j(j+1)-k(k+1)$, and where the dotted line
denotes that the two vertices are at the same point, i.e. the line
between the two vertices carries a holonomy equal to the identity.

Using the rearrangement, we get,
\begin{equation}
\Delta_{ab}(\pi_o^x) E(\Gamma,\kappa) = -\sum_{e_C^D \in \Gamma} 
\kappa \sum_{l=|j-k|}^{j+k} (2 l +1 ) \rho_l(j,k) \epsilon_{abc}
\int_{e_C^D} dy^c \delta^3(x-y) \left<
\raisebox{-10mm}{\psfig{figure=idayvuel.eps,height=20mm}}
\right>{}_{\raisebox{-3mm}{$\scriptstyle CS$}}
,\label{idayvuel}
\end{equation}
and then using the basic recoupling property of $SU(2)$ spin networks,
\begin{equation}
\raisebox{-10mm}{\psfig{figure=paral.eps,height=20mm}} = 
\sum_{m=0}^{2j} (-1)^{2j} (2 m +1 ) 
\raisebox{-10mm}{\psfig{figure=yy.eps,height=20mm}},
\end{equation}
we get, 
\begin{equation}
\Delta_{ab}(\pi_o^x) E(\Gamma,\kappa) = 
-\sum_{e_C^D \in \Gamma} \kappa \sum_{l=|j-k|}^{j+k}
\sum_{m=0}^{2j}
(-1)^{2j} (2 m +1) (2 l+1) \rho_l \epsilon_{abc} \int_{e_C^D} dy^c
\delta^3(x-y) \left<
\raisebox{-10mm}{\psfig{figure=idayvuel2.eps,height=20mm}}
\right>{}_{\raisebox{-3mm}{$\scriptstyle CS$}}
\label{idayvuel2}
\end{equation}
and applying recoupling in the dotted circle in (\ref{idayvuel2}) and
using the 6j symbols, one gets the final expression,
\begin{equation}
\Delta_{ab}(\pi_o^x) E(\Gamma,\kappa) = 
\sum_{e_C^D \in \Gamma} \sum_{m=0}^{2j} -\kappa
\epsilon_{abc}
\int_{e_C^D} dy^c \delta^3(x-y) \lambda^{\pm}_m(j,k)\left<
\raisebox{-10mm}{\psfig{figure=idayvuel3.eps,height=20mm}}
\right>{}_{\raisebox{-3mm}{$\scriptstyle CS$}},\label{resfinal}
\end{equation}
where,
\begin{equation}
\lambda^\pm_m(j,k) =(-1)^{j\mp k} \sum_{l=|j-k|}^{j+k} (-1)^{l-(m\mp
m)/2} (2 m+1)(2l+1) \rho_l \left\{
\begin{array}{ccc}
j&j&m\\k&k&l\end{array}\right\}.
\end{equation}

So the end result is that the loop derivative acting on the invariant
$E(\Gamma,\kappa)$ is nonvanishing only if the endpoint of the path
$\pi_o^x$ of the loop derivative falls upon one of the lines of the
spin network. The end result is, up to a factor, proportional to the
invariant $E(\Gamma',\kappa)$, where $\Gamma'$ is a new graph obtained
by adding the path $\pi_o^x$ to the original graph $\Gamma$. Notice
that the action of the loop derivative is covariant under
diffeomorphisms, in the sense that a diffeomorphism would shift both
the graph $\Gamma$ and the path $\pi_o^x$ and therefore would act as a
diffeomorphism on the graph $\Gamma'$.  It is also worthwhile noticing
that the general form of equation (\ref{resfinal}) is true for any
gauge group, the only differences would appear in the recoupling
coefficients $\lambda$ and the corresponding irreducible
representations associated with the graph. This emphasises the
geometrical nature of the action of the loop derivative and may open
possibilities for generalizing this construction for invariants that
do not necessarily arise from $SU(2)$ groups. The importance of this
is that it appears that Vassiliev invariants for spin nets might arise
as linear combinations of products of 
``primitive'' invariants associated only
with the topological embedding of the graph of the spin net, times
some group-dependent factors that contain information about the
valences of each line in the spin net. The loop derivative acts on
such objects by ignoring the group prefactors and acting on the
factors depending on the embedding of the spin net diagrams. In
particular, if one acts on $E(\Gamma,0)$, since it does not have
information about the embedding, the loop derivative automatically
gives zero. 

The loop derivative as defined in this paper 
has several appealing properties that other differential operators in
loop space do not have (see \cite{GaPubook} for more details). One
property we wish to emphasize is that the loop derivative satisfies
Leibnitz' rule. It acts on a product of functions exactly like an
ordinary derivative. This allows, when evaluating operators on
products of knots (as for instance when we extract the frame dependent
prefactor and the $E(\Gamma,0)$), to perform explicit calculations. 

For instance, we can compute the action of the loop derivative on any
Vassiliev invariant of type $v_n$, by computing the logarithmic
derivative of $E(\Gamma,\kappa)$. The final result is,
\begin{equation}
\Delta_{ab}(\pi_o^x) v_n(\Gamma) = -
\sum_{e_j \in \Gamma} \sum_{m=0}^{2j} \lambda_m \epsilon_{abc} \int_{e_j} 
dy^c \delta^3(x-y) {E(\Gamma_m,0)\over E(\Gamma,0)} {1\over (n-1)!} 
\left[{d^{n-1} \over d
\kappa^{n-1}} 
\exp\left(\sum_r \left(v_r(\Gamma_m)- v_r(\Gamma)\right)
\kappa^r\right)\right]_{\kappa=0}\label{loopdonvas}
\end{equation}
and the action is nonvanishing only if $o,x$ fall on two different lines
of $\Gamma$ and the spin net $\Gamma_m$ is obtained by taking $\Gamma$
and adding to it the line $\pi_o^x$ with valence $m$.

\section{The diffeomorphism constraint}
\subsection{The operator}

In this section we will introduce the diffeomorphism constraint as a
differential operator in loop space written in terms of the loop
derivative. We will also show that the invariants we introduced in the
previous section will be either annihilated by the constraint in the
case of ambient isotopy invariants, or will transform appropriately in
the case of regular isotopic invariants. 

Let us consider the diffeomorphism constraint of quantum gravity
written in terms of Ashtekar's new variables,
\begin{equation}
\hat{C}(\vec{N}) \Psi[A] = \int d^3x N(x)^a \hat{\tilde{E^b_i}} (x)
\hat{F}_{ab}^i(x) \Psi[A].
\end{equation}
Because this expression involves the product of two operators at the
same point, in general we need to regularize it, which we do via a
point splitting function $\lim_{\epsilon\rightarrow 0} f_\epsilon(x,y)
= \delta^3(x,y)$,
\begin{equation}
\hat{C}(\vec{N}) \Psi[A] = \lim_{\epsilon\rightarrow 0} 
\int d^3 x \int d^3 y \, f_\epsilon(x,y) N^a(x) 
\hat{\tilde{E}}^b_i(x)
[U(\pi_x^y) \hat{F}_{ab}(y) U^{-1}(\pi_y^x)]^i \Psi[A]\label{diffcon}
\end{equation}
where in order to preserve gauge invariance we have connected the
$F_{ab}$ and $\tilde{E}$ operators with holonomies along a path $\pi$
going from $x$ to $y$.

The above operator, when acting on a Wilson net, can be rewritten in
terms of the loop derivative, as has been discussed in the context of
loops in the past \cite{GaPubook}. The explicit expression is,
\begin{equation}
\hat{C}(\vec{N}) W(\Gamma,A) = \lim_{\epsilon\rightarrow 0} 
\int d^3 x \, f_\epsilon(x,y) 
\sum_{e_A^B \in \Gamma} \int_{e_A^B}  dy^b N^a(y) 
\Delta_{ba}(\pi_y^x) W(\Gamma,A). \label{gendif}
\end{equation}
One can explicitly check that the action of this operator is a
diffeomorphism, provided that the connection $A$ is smooth. In the
limit in which the regulator is removed, the path $\pi$ shrinks to a
point and one ends with a loop with an infinitesimal closed loop
attached at the point $x$. The addition of this closed loop is
tantamount to displacing infinitesimally the line of the original loop
at the point $x$.

We will now assume that the generator of diffeomorphisms has in
general the form given by equation (\ref{gendif}) in terms of the loop
derivative and we will show that when acting with it on the invariants
from Chern--Simons we get the correct result. This result is
nontrivial, since in the path integral that defines the invariant
there are contributions from distributional connections. 

\subsection{Action on regular isotopic invariants}

To discuss in a cleaner fashion the action of the diffeomorphism
operator on the invariants, we will consider diffeomorphisms in which
the vector $\vec{N}$ has compact support. This will allow us to focus
on the action of diffeomorphisms on individual edges, on vertices,
etc, each at a single time. It is immediate that a generic situation
can be analyzed combining all results we will derive. Let us start
with the action at an individual line, (we define $\hat{C}(\vec{N}) =
\lim_{\epsilon \rightarrow 0} \hat{C}_\epsilon(\vec{N})$)
\begin{eqnarray}
\hat{C}_\epsilon(\vec{N})\left<
\raisebox{-10mm}{\psfig{figure=vertj.eps,height=20mm}}
\right>{}_{\raisebox{-3mm}{$\scriptstyle CS$}}  &=& 
\kappa \sum_{m=0}^{2j} \lambda^-_m(j,j) \int dy^b \int dz^c
f_\epsilon(z,y) \epsilon_{abc} N^a(z) \times\\
&&{1 \over 2} [
\Theta(z-y) 
\left<\raisebox{-10mm}{\psfig{figure=vertbubu.eps,height=20mm}}
\right>{}_{\raisebox{-3mm}{$\scriptstyle CS$}} 
+
\Theta(y-z) 
\left<\raisebox{-10mm}{\psfig{figure=vertbubd.eps,height=20mm}}
\right>{}_{\raisebox{-3mm}{$\scriptstyle CS$}} ],\nonumber
\end{eqnarray}
where $\Theta$ are Heaviside functions that order points along the line
of the spin net we are considering. One can easily define them
introducing a parameterization.  We now use a recoupling identity,
\begin{equation}
\left<\raisebox{-10mm}{\psfig{figure=vertbubu.eps,height=20mm}}
\right>{}_{\raisebox{-3mm}{$\scriptstyle CS$}}  =
{(-1)^{2j} \over (2 j+1)} \{j,j,m\} \left<
\raisebox{-10mm}{\psfig{figure=vertjbaj.eps,height=20mm}}
\right>{}_{\raisebox{-3mm}{$\scriptstyle CS$}}   
\end{equation}
where $\{j,j,m\}$ is the $3j$ symbol that is zero unless the $j$ and
$m$ values satisfy the triangular relation at the vertices. 
We now remove the regulator, assuming the following regularization
function, 
\begin{equation}
f_\epsilon(y,z) = {3 \over 4 \pi \epsilon^3} \Theta(\epsilon -|z-y|).
\end{equation}
Evaluating the integral, one gets,
\begin{equation}
\hat{C}(\vec{N})\left<
\raisebox{-10mm}{\psfig{figure=vertj.eps,height=20mm}}
\right>{}_{\raisebox{-3mm}{$\scriptstyle CS$}}  = 
{\kappa \over 8 \pi}
\sum_{m=0}^{2j} \lambda^-_m(j,j) {(-1)^{2j}\over (2j+1)}
\int_0^1 ds\, \epsilon_{abc} {\dot{N}(s)^a
\dot{y}(s)^c \ddot{y}(s)^b
\over |\dot{y}(s)|^3}
\left<\raisebox{-10mm}{\psfig{figure=vertj.eps,height=20mm}}
\right>{}_{\raisebox{-3mm}{$\scriptstyle CS$}} 
\end{equation}
where we have introduced a parameterization such that $dy^a =
\dot{y}(s)^a ds$ and dots refer to total derivatives with respect to
$s$. We can summarize the above result by saying that, for
diffeomorphisms of compact support acting  on lines of the spin net,
we have that,
\begin{equation}
\hat{C}(\vec{N}) E(\Gamma,\kappa) = \kappa \mu(j) w(\vec{N}) E(\Gamma,\kappa)
\label{diffone}
\end{equation}
where $\mu(j)=\sum_{m=0}^{2j} \lambda^-_m(j,j) {(-1)^{2j}\over (2j+1)}$
is a recoupling factor, that depends on the weight of the line in
question, and $w(\vec{N})$ is the writhe introduced in the line by the
action of the diffeomorphism along the vector $\vec{N}$. We therefore 
see that the diffeomorphism has a well defined limit. The result is
dependent on a background metric through the writhe, as expected for a
regular istopic invariant. Notice that the result decomposes as a
product of a factor depending entirely on the ``coloring'' of the spin
net and another factor depending only on the embedding of the net in
three dimensions. What we have therefore recovered is a formula that 
contains the information about the non-invariance of $E(\Gamma,k)$
under the addition of twists, given by  the skein
relation (\ref{twist}). This skein relation could be reobtained
exactly by exponentiating the action of the loop derivative, as
discussed in \cite{GaPucmp}, but we will not repeat the calculation
here for brevity.

Let us now consider the action of the diffeomorphism at an
intersection. Here we get, in addition to the contributions we
discussed for a single line, a contribution when the end of the path
of the loop derivative intersects one of the lines entering the
intersection. We have,
\begin{equation}
\hat{C}(\vec{N})\left<
\raisebox{-10mm}{\psfig{figure=y+.eps,height=20mm}}
\right>{}_{\raisebox{-3mm}{$\scriptstyle CS$}}  = \kappa
\sum_{\alpha\ne\beta =1}^3 \sum_{m=0}^{2 j_\alpha}
\lambda^-_m(j_\alpha,j_\beta) \int dy^b \int dz^c f_\epsilon(y,z)
N^a(z) \epsilon_{abc} 
\left<\raisebox{-10mm}{\psfig{figure=ybuba.eps,height=20mm}}\right>
{}_{\raisebox{-3mm}{$\scriptstyle CS$}},
\end{equation}
and using the recoupling identity,
\begin{equation}
\left<\raisebox{-10mm}{\psfig{figure=ybubano.eps,height=20mm}} 
\right>{}_{\raisebox{-3mm}{$\scriptstyle CS$}}  
=
{ (-1)^{m+j_\gamma-j_\alpha-j_\beta} 
\left\{
\begin{array}{ccc}
j_\alpha & j_\beta & m\\j_\beta & j_\alpha & j_\gamma
\end{array}\right\} \over \Theta(j_\alpha,j_\beta,j_\gamma)}
\left<\raisebox{-10mm}{\psfig{figure=yno.eps,height=20mm}}
\right>{}_{\raisebox{-3mm}{$\scriptstyle CS$}}  
,
\end{equation}
and removing the regulator as before, we get,
\begin{equation}
\hat{C}_\epsilon(\vec{N})\left<
\raisebox{-10mm}{\psfig{figure=y+.eps,height=20mm}}
\right>{}_{\raisebox{-3mm}{$\scriptstyle CS$}}  = \kappa
\sum_{\alpha=1}^2 \sum_{\beta=\alpha+1}^3 \Omega_m(j_\alpha,j_\beta,j_\gamma)
\epsilon_{abc} \dot{N}^a(0) \dot{y}^b_\alpha(0) \dot{y}^c_\beta(0)
\left<\raisebox{-10mm}{\psfig{figure=y+.eps,height=20mm}}
\right>{}_{\raisebox{-3mm}{$\scriptstyle CS$}}  
\end{equation}
where,
\begin{equation}
\Omega_m(j,k,l)= Q (-1)^{m+l-j-k} 
\left\{
\begin{array}{ccc}
j & k & m\\ k & j & l
\end{array}\right\} \lambda^-_m(j,k)
\end{equation}
with $Q$ a constant given by,
\begin{equation}
Q = {1\over 4\pi}
\int_0^{\pi/2} d \phi  { (\sin \phi - \cos \phi) \over \sqrt{
|\dot{\vec{y}}_\alpha(0)|^2 \cos^2 \phi + 
|\dot{\vec{y}}_\beta(0)|^2 \sin^2 \phi -
2 |\dot{\vec{y}}_\alpha(0)\cdot \dot{\vec{y}}_\beta(0)| \cos \phi \sin \phi}}.
\end{equation}
In the above equations we have assumed a parameterization with
parameter equal to zero at the intersection. The geometrical meaning
of this expression is that it vanishes if the volume spanned by the
two tangents at the intersection and the derivative of the vector
$\vec{N}$ along one of the lines entering the intersection
vanishes. If this volume is nonzero, we start to reconstruct
infinitesimally the twist at the intersection covered by the skein
relation (\ref{ytwist}). Again, one could recover it exactly
exponentiating the action of the loop derivative as described in
\cite{GaPucmp}.

\subsection{Action on ambient isotopic invariants}

Here we will show that the diffeomorphism operator we introduced
vanishes when acting on the ambient isotopic invariants, concretely
$P(\Gamma,\kappa)$. Therefore, it annihilates every Vassiliev
invariant. In order to compute the action, let us recall equation
(\ref{descomp}),
\begin{equation}
E(\Gamma,\kappa) = E(\Gamma,0) (1+v_1(\Gamma) \kappa +(2 v_2(\Gamma) +
v_1(\Gamma)^2)
{\kappa^2\over 2}+\cdots)
\end{equation}
and analyze the action of the diffeomorphism on $E(\Gamma,\kappa)$
order by order in $\kappa$. To order zero we get, 
\begin{equation}
\hat{C}(\vec{N}) E(\Gamma,0) =0
\end{equation}
which is correct, since $E(\Gamma,0)$ does not depend on the embedding
of the spin net, just on its coloring, and it is immediately
annihilated by the loop derivative. At the next order, we have,
\begin{equation}
\hat{C}(\vec{N}) v_1(\Gamma) = \mu(j) w(\vec{N}) \label{diffonv1}
\end{equation} 
where we have assumed the diffeomorphism as acting on an isolated line
of valence $j$, a similar formula holds for the intersections with a
different coloring weight. As before $w(\vec{N})$ is the writhe
induced in the line by the vector field $\vec{N}$. If we now take
advantage of the fact that the diffeomorphism operator satisfies
Leibnitz' rule, we see that combining (\ref{diffone}) and
(\ref{diffonv1}) we get, 
\begin{equation}
\hat{C}(\vec{N}) \exp(-\kappa v_1(\gamma)) E(\Gamma,\kappa) =0,
\end{equation}
and therefore we see that indeed the diffeomorphism constraint
annihilates all Vassiliev invariants, namely,
\begin{equation}
\hat{C}(\vec{N}) P(\Gamma,\kappa)  = 
\hat{C}(\vec{N}) \exp\left(\sum_{n=2}^\infty v_n(\Gamma) \kappa^n\right)=0.
\end{equation}

Since the diffeomorphism constraint annihilates all Vassiliev
invariants, it also annihilates any function of the Vassiliev
invariants. As we mentioned before, there are indications that
Vassiliev invariants may constitute a basis of diffeomorphism
invariant functions of loops. What we have shown here is that the
definition of the loop derivative we introduced in the spin network
context is compatible with that fact, namely it naturally leads to a
diffeomorphism constraint that annihilates explicitly all Vassiliev
invariants. 

\section{Towards a Hamiltonian constraint}

Let us now consider the Hamiltonian constraint of quantum
gravity, possibly with a cosmological constant $\Lambda$,
\begin{equation}
\hat{H}(\subtilde{M}) = \int d^3x \subtilde{M} 
\left[\epsilon^{ijk} \hat{\tilde{E}}^a_i 
\hat{\tilde{E}}^b_j \hat{F}_{ab}^k  + {\Lambda \over 6}
\epsilon^{ijk} \hat{\tilde{E}}^a_i 
\hat{\tilde{E}}^b_j \hat{\tilde{E}}^c_k \epsilon_{abc}\right].
\end{equation}
This expression corresponds to the ``doubly densitized'' operator. We
chose to do this since we have experience with this operator from the
context of loops \cite{GaPubook} and it is easily expressed in terms
of the loop derivative. We will later suggest how one could transform
this expression into a single densitized version.
We need to regularize the operator, which again we do through a
point splitting. There are many options as to how to exactly point
split. The way we do it, for the $\Lambda=0$ piece, is,
\begin{equation}
g_{\epsilon,\epsilon',\epsilon''}
(u,v,w,x) H(u,v,w) \equiv 
g_{\epsilon,\epsilon',\epsilon''}
(u,v,w,x) \epsilon^{ijk} \hat{\tilde{E}}^a_i(u)
\hat{\tilde{E}}^b_j(v) [U (\pi_u^w) F_{ab}(w) U (\pi_w^u)]^k.
\end{equation}
So we have chosen to join the $F_{ab}$ with one of the $E$'s via a
holonomy. There are many other possibilities (for instance, 
joining all operators via holonomies, which would yield a gauge
invariant regularization) and they all yield the same classical expression
in the the limit
$\epsilon\rightarrow 0$. A possible regulating function is defined as,
\begin{equation}
g_{\epsilon,\epsilon',\epsilon''}(u,v,w,x) = f_\epsilon(x,u)
f_\epsilon(x,{u+v\over 2}) f_\epsilon(x,{u+v+w\over 3})
\end{equation}
with $f$ defined as in the diffeomorphism constraint. This regulating 
function is the same as the one usually considered for the
``Ashtekar--Lewandowski'' volume operator,
\cite{AsLe,Thvo}. It is worthwhile pointing
out that for the following discussion, this issue is not important,
since we will discuss results that do not require to remove the regulator. 

To determine the action of the Hamiltonian in terms of spin networks,
we consider the action of the operator we have just defined on a
Wilson net. As is well known, this operator only acts at intersections
of the net. At regular points it gives rise to ``acceleration'' terms,
which we will omit since they vanish on diffeomorphism invariant
states \cite{BrPu,GaPubook}. The nonvanishing action comes from the
triads acting at two points on different strands $i,k$ entering an
intersection $V$,  
\begin{eqnarray}
H(u,v,w) W\left(
\raisebox{-10mm}{\psfig{figure=yjk.eps,height=20mm}},A\right) 
&=& -{i \over 2} \sum_{i \ne k\in V}
\sum_{m=|i-k|}^{i+k} (2m+1) \rho_m(i,k) \int_{e_i} dy^a \delta^3(u-y) 
\int_{e_k} dz^b \delta^3(v-z) \times \label{hamspin}\\
&&\left[
\Delta_{ab}(\pi_{y+}^w) -\Delta_{ab}(\pi_{y-}^w) \right]
W\left(
\raisebox{-15mm}{\psfig{figure=yham.eps,height=30mm}},A\right)\nonumber
\end{eqnarray}
where in the last diagram all the structure occurs ``at a point'' and
points $y\pm$ are identified. 

Having the Hamiltonian in terms of the loop derivative allows to
evaluate its action on any of the invariants we have been discussing.
Let us start by showing that the Hamiltonian constraint with a
cosmological constant term annihilates the invariant
$E(\Gamma,\kappa)$. In order to do this, we need to derive an
expression for the operator $\hat{\rm det} q$, $q$ being the spatial
metric in terms of spin nets. The construction completely parallels
that of the Hamiltonian constraint, the operator $\hat{\rm det} q$ is
obtained by replacing $F_{ab}^i$ in the Hamiltonian with
$\epsilon_{abc}\tilde{E}^c_i$. In order to be possible to have
a cancellation with the Hamiltonian, one has to regularize both
operators in a similar manner. So we choose,
\begin{equation}
g_{\epsilon,\epsilon',\epsilon''}
(u,v,w,x) \hat{{\rm det}}(u,v,w) \equiv {i \over 3!}
g_{\epsilon,\epsilon',\epsilon''}
(u,v,w,x) \epsilon^{ijk} \hat{\tilde{E}}^a_i(u)
\hat{\tilde{E}}^b_j(v) [U (\pi_u^w) \tilde{E}^c(w) U (\pi_w^u)]^k
\epsilon_{abc}.
\end{equation}
The point-splitting (i.e. where one decides to insert each
holonomy) could have been made in a different way
(e.g. \cite{AsLe,Th}), but one can show ---using recoupling theory---
that inserting additional holonomies leads in the limit when one
removes the regulator to the same result for a given regulating function.

The end result for the operator is, 
\begin{eqnarray}
\hat{{\rm det} g}(u,v,w) 
\Psi\left(\raisebox{-10mm}{\psfig{figure=yjk.eps,height=20mm}}\right) &=& 
{1 \over 24} \sum_{i,k,l} \sum_{m,n} (2 m+1) \rho_m(i,k) 
\lambda^-_n(i,l) \epsilon_{abc} \times \label{detspin}\\
&&\int_{e_i} dy^a \delta^3(y,u) 
\int_{e_k} dz^b \delta^3(z,v) 
\int_{e_l} dt^c \delta^3(t,w) \times\nonumber\\
&&\left[ 
\Psi\left(\raisebox{-12mm}{\psfig{figure=det3a.eps,height=25mm}}\right)-
\Psi\left(\raisebox{-12mm}{\psfig{figure=det3b.eps,height=25mm}}\right)
\right].\nonumber
\end{eqnarray}
The vertex on which the determinant acts can be of any order, if it is
higher than trivalent, the determinant is a sum of the above
contribution evaluated for all possible triplets of lines. 
All linkages in the figure are
assumed to be happening ``at a point'', so recoupling theory can be
applied.  An interesting observation is that if the vertex is
trivalent, simple recoupling identities show that both contributions
are equal and therefore the determinant of the metric vanishes on such
vertices (as one would have suspected, given that the volume operator
vanishes too).

Using the definition of the above operators, one can start obtaining
results. The first thing we notice is that,
\begin{equation}
g_{\epsilon,\epsilon',\epsilon''}
(u,v,w,x) \hat{H}(u,v,w) E(\Gamma,\kappa) = 
12 \kappa i
g_{\epsilon,\epsilon',\epsilon''}
(u,v,w,x) \hat{{\rm det} g}(u,v,w) E(\Gamma,\kappa)
\end{equation}
and we therefore see that the generalization of the Kauffman bracket
to spin networks $E(\Gamma,\kappa)$ is a solution of the Hamiltonian
constraint of quantum gravity with a cosmological constant. This is 
straightforward to see by considering the action of
(\ref{hamspin}) and making explicit the action of the loop derivative,
which adds a line to the diagram in such a way that one ends with the
same diagrams as in (\ref{detspin}).

Another interesting result is to notice that the Hamiltonian vanishes
when acting on Vassiliev invariants at trivalent vertices. This can be
straightforwardly seen by recalling the action of the loop derivative
on a Vassiliev (\ref{loopdonvas}), and noting that at a triple
intersection, using recoupling, $v_p(\Gamma_j)=v_p(\Gamma)$. We are
allowed to use recoupling because the loop derivative acts by adding a
line ``at a point'' when evaluated in the action of the Hamiltonian
constraint. Therefore the Hamiltonian constraint vanishes at
trivalent vertices.

Another important result concerning the action of the Hamiltonian on
Vassiliev invariants can again be concluded from studying the explicit
action of the loop derivative (\ref{loopdonvas}). Combining this
expression with (\ref{hamspin}) one immediately concludes that the
action of the ``doubly-densitized'' Hamiltonian we are considering
here is proportional to the ``regularized volume'' (strictly speaking
it is a volume squared) spanned by three of the
tangents entering at an intersection \cite{Br},
\begin{equation}
\hat{\rm Vol}^{ijk}_{\rm reg}(x,\epsilon,\epsilon',\epsilon'') 
= \epsilon_{abc} 
\int_{e_i} dy^a 
\int_{e_j} dz^a 
\int_{e_k} dw^a 
g_{\epsilon,\epsilon',\epsilon''}
(x,y,z,w).
\end{equation}
Concretely, acting on a four-valent vertex labelled by $J$, 
the action of the Hamiltonian is,
\begin{equation}
g_{\epsilon,\epsilon',\epsilon''}(x,u,v,w) \hat{H}_{\rm double}(u,v,w)
 v_n(
\left(\raisebox{-5mm}{\psfig{figure=wishj.eps,height=10mm}}\right)
= \sum_I \sum_{i,j,k \in {\rm Vertex}}
\hat{\rm Vol}^{ijk}_{\rm reg}(x,\epsilon,\epsilon',\epsilon'') 
 C^{IJ}V_{n-1}
\left(\raisebox{-5mm}{\psfig{figure=wishi.eps,height=10mm}}\right)
\end{equation}
with $C^{IJ}$ is a proportionality factor depending on the valences 
of the strands entering the intersection. Similarly, the ranges of the
sum in $I$ is determined by the type of intersection. The invariant
$V_{n-1}$ is a non-primitive Vassiliev invariant of order $n-1$, it
will involve sums of products of primitive Vassiliev invariants of 
lower orders.

This proportionality opens the attractive possibility of defining a
``singly densitized'' Hamiltonian by ``dividing by the regularized
volume''. Having a singly densitized Hamiltonian can potentially lead
to a much better defined operator, since its action could possibly be
cast in a background-independent manner, as proportional to a Dirac
delta.  If this construction works, it might therefore be possible to
cast the action of the Hamiltonian constraint in terms of an expression
along the lines of,
\begin{equation}
H_{\rm single}(x) v_n
\left(\raisebox{-5mm}{\psfig{figure=wishj.eps,height=10mm}}\right)
= \delta(x-{\rm Vertex}) \sum_I
d^{IJ} V_{n-1}
\left(\raisebox{-5mm}{\psfig{figure=wishi.eps,height=10mm}}\right)
\end{equation}
with $d^{IJ}$ another constant factor. Having an operator with such
a simple action is not only remarkable, but will open the possibility
of further consistency checks, like computing the constraint
algebra. All these issues are currently under study and evidently
further work is needed to complete this program. The computation can
be done ``on shell'' by showing that the Hamiltonian commutes on
Vassiliev invariants using the expressions introduced above. It could
also be done ``off shell'' since one has an expression for the
Hamiltonian acting on any function that is loop differentiable, and
therefore verify commutation relations with the diffeomorphism
constraint.

\section{Conclusions}

The doubly-densitized Rovelli--Smolin Hamiltonian has been studied,
written in terms of the loop derivative, since the early days of the
new variable formulation in terms of loops. The early work suffered
from three main drawbacks: a) the loop derivative was not a well
defined operator on diffeomorphism invariant states; b) the quantum
states were difficult to construct and the action of the constraint to
check explicitly due to the presence of the Mandelstam identities and
c) the Hamiltonian was regularization dependent, potentially having
problems at the time of computing the quantum commutator algebra of
constraints. In this work, largely by going to the language of spin
networks, we have made progress in all three points. First of all, we
were able to introduce a loop derivative that is well defined when
acting on Vassiliev invariants. The loop derivative leads, while input
into the usual definition, to a well defined diffeomorphism
constraint, with exactly the kind of action on Vassiliev invariants
and its regular-isotopic counterparts, as one would expect.  The
presence of this reasonable-looking loop derivative operator allows us
for the first time to actually have a version of the Rovelli--Smolin
Hamiltonian written in terms of the loop derivative that actually is a
well defined operator on knots. We have checked that several of the
formal results obtained in the loop language still go through with the
new operator. Better yet, since all the results are in terms of spin
network states, we need not concern ourselves anymore with the
Mandelstam identities. Finally, since the Hamiltonian has a relatively
simple action on Vassiliev invariants, closely related to the volume
operator, the possibility is opened of defining a single-densitized
Hamiltonian, just by dividing by the operator representing the
determinant of the metric. This latter possibility still needs to be
studied in detail. However, preliminary results appear as
promising. The single densitized operator appears to have a very
natural action on the space of knots. This opens the hope that a
consistent set of constraints, satisfying the appropriate commutator
algebra, might be found. One could think of computing the commutator 
algebra off-shell, very much along the same lines as in the formal
computations of \cite{GaGaPu}, except that now the operators involved
are also well defined on-shell (i.e. when the wavefunctions are knot
invariants) . The fact that all this occurs while working
on the space of Vassiliev knot invariants based on spin networks
suggests that this might be an appropriate setting for the discussion
of canonical quantum gravity.

\acknowledgements

We wish to thank Laurent Freidel for discussions.
This work was supported in part by grants NSF-INT-9406269,
NSF-PHY-9423950, research funds of the Pennsylvania State University,
the Eberly Family research fund at PSU and PSU's Office for Minority
Faculty development. JP acknowledges support of the Alfred P. Sloan
foundation through a fellowship. We acknowledge support of Conicyt
(project 49) and PEDECIBA (Uruguay).

\end{document}